\pdfoutput=1
\documentclass[10pt,conference,letterpaper]{IEEEtran}

\usepackage[noadjust]{cite}
\usepackage[T1]{fontenc}
\usepackage[utf8]{inputenc}
\usepackage{balance}
\usepackage{url}
\usepackage{subcaption}
\usepackage{rotating}
\usepackage{bigdelim}
\usepackage{multirow}
\usepackage{tabularx}
\usepackage{hhline}
\usepackage{colortbl}  %
\usepackage{dblfloatfix}  %

\newcommand{\docauthor}{Jonathan Will\IEEEauthorrefmark{1}, Lauritz Thamsen\IEEEauthorrefmark{2}, Jonathan Bader\IEEEauthorrefmark{1}, Dominik Scheinert\IEEEauthorrefmark{1}, and Odej Kao\IEEEauthorrefmark{1}}
\newcommand{\docsubject}{\IEEEauthorrefmark{1}Technische Universit\"at Berlin, Germany \hspace{6mm} \IEEEauthorrefmark{2}University of Glasgow, United Kingdom}
\newcommand{\dockeywords}{Scalable Data Analytics, Distributed Dataflows, Profiling, Resource Allocation, Cluster Management}
\newcommand{\doctitle}{Ruya: Memory-Aware Iterative Optimization of Cluster Configurations for Big Data Processing}

\PassOptionsToPackage{bookmarks=false}{hyperref}
\usepackage[pdftex]{hyperref}
\hypersetup{
    pdfborder={0 0 0},
    pdfauthor=\docauthor,
    pdftitle=\doctitle,
    pdfsubject=\docsubject,
    pdfkeywords=\dockeywords
}

\usepackage{amsmath,amssymb,amsfonts}
\usepackage{algorithmic}
\usepackage{graphicx}
\usepackage{textcomp}
\usepackage{tikz}
\usepackage{xcolor}  %

\definecolor{newgreen}{RGB}{7, 160, 25}
\newenvironment{gn}{\color{newgreen}}

\definecolor{Gray}{gray}{0.9}

\def\mycopyrightnotice{
  {\footnotesize 978-1-6654-8045-1/22/\$31.00~\copyright~2022 IEEE\\
   DOI: \href{https://doi.org/10.1109/BigData55660.2022.10020295}{https://doi.org/10.1109/BigData55660.2022.10020295}
  }
  \gdef\mycopyrightnotice{}
}
\makeatletter
  \def\ps@IEEEtitlepagestyle{%
 \def\@oddfoot{\mycopyrightnotice}
  \def\@evenfoot{}
  }%
\makeatother

\def\BibTeX{{\rm B\kern-.05em{\sc i\kern-.025em b}\kern-.08em T\kern-.1667em\lower.7ex\hbox{E}\kern-.125emX}}

\begin{document}

\title{\doctitle}

\author{%
\IEEEauthorblockN{\docauthor}
\IEEEauthorblockA{\docsubject\\
\{will, jonathan.bader, dominik.scheinert, odej.kao\}@tu-berlin.de \hspace{6mm} lauritz.thamsen@glasgow.ac.uk
}}

\maketitle

\begin{abstract}
Selecting appropriate computational resources for data processing jobs on large clusters is difficult, even for expert users like data engineers.
Inadequate choices can result in vastly increased costs, without significantly improving performance.
One crucial aspect of selecting an efficient resource configuration is avoiding memory bottlenecks.
By knowing the required memory of a job in advance, the search space for an optimal resource configuration can be greatly reduced.

Therefore, we present \textit{Ruya}, a method for memory-aware optimization of data processing cluster configurations based on iteratively exploring a narrowed-down search space.
First, we perform job profiling runs with small samples of the dataset on just a single machine to model the job's memory usage patterns.
Second, we prioritize cluster configurations with a suitable amount of total memory and within this reduced search space, we iteratively search for the best cluster configuration with Bayesian optimization.
This search process stops once it converges on a configuration that is believed to be optimal for the given job.
In our evaluation on a dataset with 1031 Spark and Hadoop jobs, we see a reduction of search iterations to find an optimal configuration by around half, compared to the baseline.%
\vspace{-2mm}

\end{abstract}

\IEEEpeerreviewmaketitle


\section{Introduction}\label{sec:INTRO}
Distributed dataflow systems like Apache Flink~\cite{flink} and Apache Spark~\cite{spark} enable users from different domains such as public infrastructure monitoring, earth observation, or bioinformatics to develop scalable data-parallel programs~\cite{geldenhuys2021dependable, bader2022lotaru, bader2022reshi}.
They reduce the need to implement parallelism and fault tolerance and, therefore, simplifying organizations' big data analysis processes.

However, it is often not straightforward to select resources and configure clusters for efficiently executing such programs~\cite{lama2012aroma, rajan2016perforator}.
To avoid resource bottlenecks and ensure performance expectations are met, users typically overprovision resources, leading to unnecessarily high costs and low resource utilization~\cite{yang2013bubble,liu2011measurement,delimitrou2014quasar,lin2013scaling}.
These issues are amplified when using larger cluster setups.
For instance, suboptimal resource configurations can increase costs tenfold when using public clouds~\cite{cherrypick, hsu2018arrow}.
Recent works~\cite{al2022juggler, al2022blink}, including our own~\cite{will2022get, will2020towards}, have noted that one of the largest cost factors is inadequate memory allocation.
Examples are iterative machine learning jobs, which rely on the dataset fitting into the combined cluster memory.
Not having enough memory then results in repeated disk read operations and high cost, while allocating more memory than necessary yields limited benefit, but also adds to the cost.
Meanwhile, a shortage or an over-abundance of other resources, like CPU often has a less significant impact on the resource-efficiency and therefore on the cost-efficiency of jobs. %

Inquiries by large companies such as Alibaba and Microsoft found that around 40\% to 65\% of data analytics jobs in their clusters are recurring~\cite{jyothi2016morpheus,wang2020grosbeak}.
Many related approaches for configuring data processing cluster resources make use of this recurrence.
State-of-the-art methods iteratively search for an optimal configuration, usually determined by the lowest total execution cost~\cite{cherrypick, hsu2018micky, hsu2018arrow, fekry2020accelerating, he2019statistics}.
Here, new configurations to be tried are selected through Bayesian optimization until it is assumed that further search will not lead to a large enough improvement to justify the additional search cost.
However, these approaches ignore the relationship between input dataset sizes and cluster memory requirements.
This makes the approaches prone to memory bottlenecks and thereby runs the risk of moving towards local optima, which slows down or even impedes finding the optimal cluster configuration.
Further, this shortcoming results in a lack of transferability of the determined optimal configurations to subsequent jobs with different input sizes, which are to be expected, even in recurring jobs as data changes or grows.

In this paper, we present \emph{Ruya}, a method for accelerating the Bayesian-optimized iterative search for the best cluster resources by incorporating information about the job's memory requirements into the search process.

As a first step, we conduct quick profiling runs on reduced hardware and dataset samples of different sizes and then monitoring the memory use of the job.
The memory requirement is then extrapolated to a job execution on the full dataset.
For this memory profiling and usage estimation, we build on and expand our previous work, Crispy~\cite{will2022get}.

In the second step, we explore the resource configuration search space, for which we use a modified version of the Bayesian optimization, as presented by Alipourfard et al.\ in {Cherry\-Pick}~\cite{cherrypick}.
Initially, we split up the search space by separating configurations catering to the job's memory need.
First, we explore the search space consisting of those prioritized configurations.
Next, the remaining configurations can be explored, until the process has reached its stopping criterion, which is the case as soon as expected performance gain does not justify the expected cost for further exploration.
In total, this leads to finding optimal and near-optimal cluster configurations quicker, reducing exploration costs and recurring execution costs.%
\vspace{3mm}

\emph{Outline}. The remainder of the paper is structured as follows.
Section~\ref{sec:BACKGROUND} provides background on allocating cluster resources to distributed dataflow jobs.
Section~\ref{sec:APPROACH} presents Ruya, our approach for efficiently navigating the resource configuration search space.
Section~\ref{sec:EVALUATION} evaluates Ruya.
Section~\ref{sec:RELATED_WORK} discusses related work.
Section~\ref{sec:CONCLUSION} summarizes and concludes this paper.

\section{Background}\label{sec:BACKGROUND}

This section explains the background on distributed dataflows running on clusters of commodity hardware and considerations about resource allocation for such jobs.

\subsection{Distributed Dataflow Jobs on Large Commodity Clusters}

Distributed dataflows are graphs of connected data-parallel operators that execute user-defined functions, typically on a set of shared-nothing commodity cluster nodes.
Such systems automatically handle failures by repeating failed operations and replacing defective nodes.
Newer distributed dataflow frameworks like Apache Spark~\cite{spark} and Apache Flink~\cite{flink} cache data in memory for faster read access, unlike the older Hadoop MapReduce~\cite{mapreduce}.
The user or the framework itself can choose strategies for data that could not fit into memory, e.g., spilling it to disk or recomputing the data from previous stages.

The nodes within the clusters that execute such programs are often virtual machines, which vary principally in their number of cores and the amount of memory per core, but some can also offer additional I/O speed or network capabilities.
In AWS, for instance, virtual machines of the \emph{c} type have less memory per core than those of the \emph{r} type, while machines of the \emph{m} type lie between those two.
Denominations like \emph{large}, \emph{xlarge}, and \emph{2xlarge} refer to the number of cores per machine.
Users of shared private clusters have to make similar considerations when allocating resources for individual jobs.

Regarding the amount of allocated resources, there is generally a trade-off between speed and cost of execution.
The different costs of individual resources like CPU cores and memory are reflected in the prices for virtual machines in public clouds, while in a private shared cluster, some resource categories can be abundant and others scarce, depending on current usage.
Hence, to increase performance efficiently, the challenge is to add the resources with the most cost-efficient performance yield.

\subsection{Memory Bottlenecks in Distributed Dataflow Jobs}

For many iterative jobs, e.g., machine learning jobs like K-Means and Page Rank, the whole dataset is read at every iteration.
To avoid repeated disk read operations or recomputations, the full dataset needs to be retained in the combined cluster memory.
Driver programs prefer data-local computations, but cached data partitions can be exchanged among nodes via the network when needed.
If the cluster does not have enough memory for perpetual in-memory processing, the job execution exhibits a significant slowdown, a \emph{memory bottleneck}.

\begin{figure}[t!]
    \centering
    \subfloat{%
      \includegraphics[width=.516\linewidth]{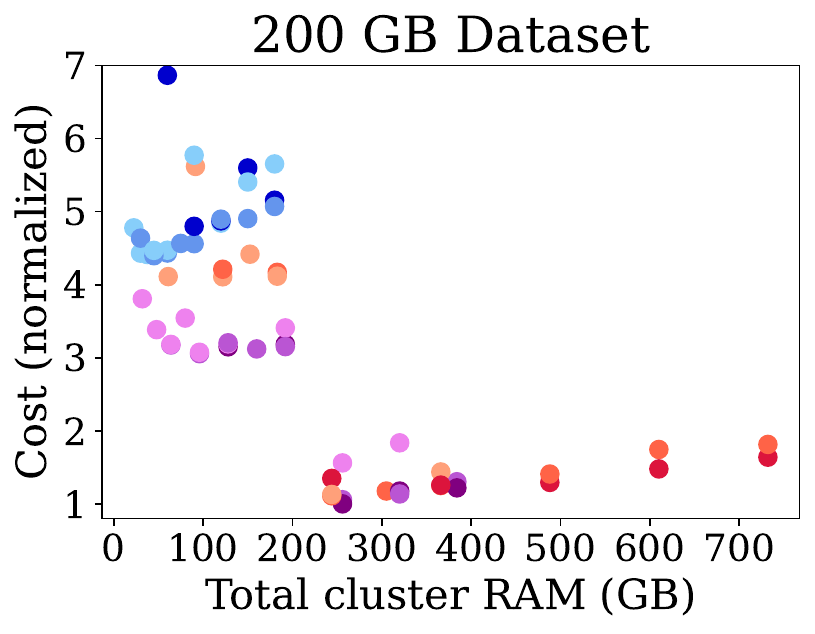}%
    }\hfill
    \subfloat{%
      \includegraphics[width=.484\linewidth]{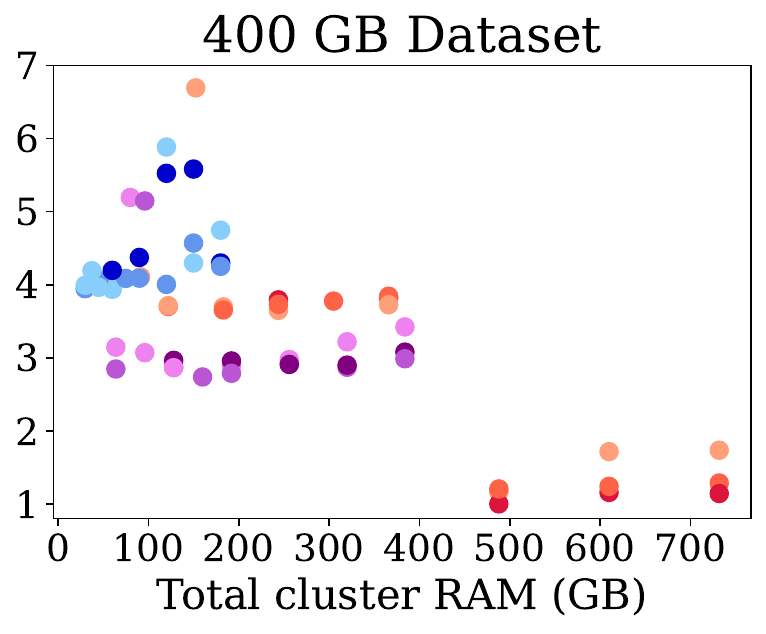}%
    }\hfill
    \subfloat{%
      \includegraphics[width=.75\linewidth]{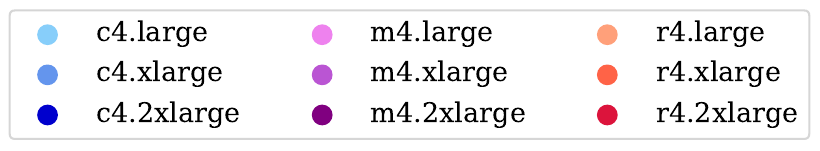}%
    }\hfill
    \caption{Total RAM vs.\ monetary cost for K-Means on Spark with different AWS machine types at different scale-outs.}\label{fig:memory_bottlenecks}
\end{figure}

In Figure~\ref{fig:memory_bottlenecks}, we see an example of memory bottlenecks for K-Means jobs on Spark.
Here, a marginal increase in total cluster memory can lead to the dataset fitting into memory,
causing drastically reduced runtime and thereby leading to a lower job execution cost.

Having enough memory for caching is crucial for preventing overhead, while additional memory is then typically ineffective in speeding up the execution.
In contrast, the performance increase for additional resources like CPU cores is more gradual.

\section{Approach}\label{sec:APPROACH}
This section presents our approach to the problem of efficiently finding optimal and near-optimal cluster configurations for recurring distributed dataflow jobs.
We first present the overall idea of the method.
Then, we explain how job profiling and memory usage estimation can accelerate the search for suitable cloud resources.

\subsection{Overview}

Utilizing knowledge about the job's memory usage patterns is at the center of our approach.
We aim to allocate enough memory to avoid unnecessary and costly repeated disk read operations and, thus, memory bottlenecks.

We propose a process for quickly and efficiently finding a good cluster resource selection, consisting of the following two steps that are visualized in Figure~\ref{fig:overview}:

\begin{figure}[htb]
    \centering
    \includegraphics[width=1\linewidth]{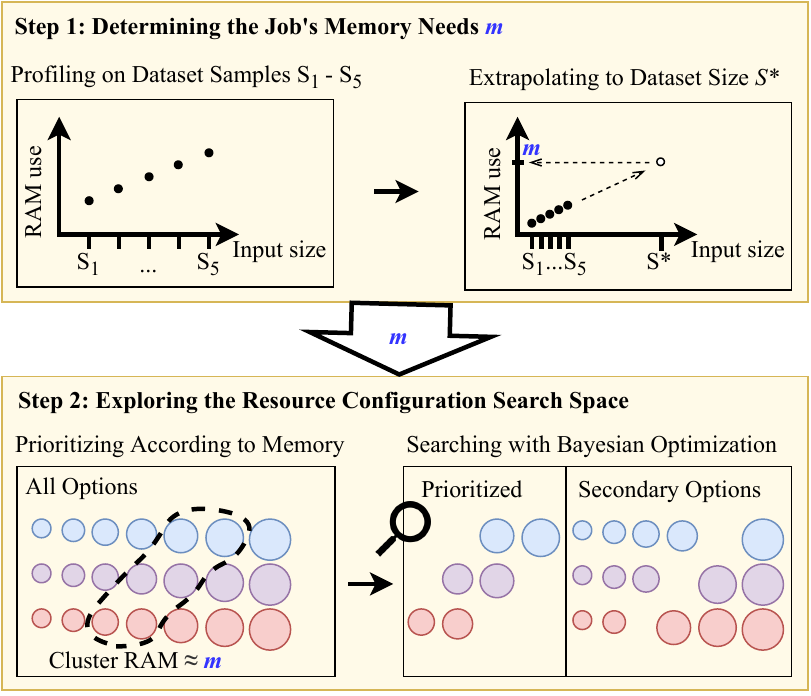}
    \caption{Overview of \emph{Ruya}, a two-step process for quickly and efficiently finding optimal cluster resource configurations.}\label{fig:overview}
\end{figure}

\textbf{1.)} The first step consists of a set of fast and resource-efficient profiling runs on reduced hardware and dataset samples of different sizes, during which we monitor the memory used by the job.
The job's memory usage is then extrapolated to a job execution on the full dataset.
This knowledge about the job's memory need is passed on to the second step.

\textbf{2.)} In the second step, we explore the configuration search space with our knowledge about the memory needs of the job that we gained in the previous step.
Initially, we split up the search space and prioritize exploring configurations that are believed to best fit the memory requirement of the job, while other configurations are only explored once the priority group has been fully tested and only until a stopping criterion is reached.

All of this allows us to adjust the selection process according to the changing size of input datasets since we know how it affects memory usage and thereby the most important factor.

\subsection{Profiling with the Crispy Resource Allocation Assistant}

In this subsection, we describe in further detail the process of conducting the profiling runs, as well as measuring and extrapolating memory usage.

For this, we make use of and expand a method that we introduced in \emph{Crispy}~\cite{will2022get}, which is an assistant for selecting the one most promising cloud configuration for a unique one-off data processing job.
Crispy does so by principally attempting to avoid memory bottlenecks and it achieves this through briefly profiling the job on reduced hardware and monitoring memory use in relation to the input dataset size.

For our current problem, we will apply its memory-aware job profiling method to accelerate finding \emph{the optimal} configuration for \emph{recurring} data processing jobs.
We conduct multiple job runs on small samples of the dataset and extrapolate the memory usage for the full dataset.
The sample sizes are chosen in a way, that they result in execution times between 30 and 300 seconds, to reach sufficiently beyond the framework's initialization phase, while also not making the profiling phase needlessly long.
This provides enough time to measure the actual memory footprint of the dataset sample.
Initially, one percent of the original dataset can be chosen and then iteratively adjusted according to match those runtime targets, i.e., if the runtime is longer than three minutes, the profiling job can be canceled and restarted with a smaller portion of that sample.
Next, four more differently sized portions of this sample are used for additional profiling runs so that the sample sizes are equally spaced and reasonably far apart to then enable modeling and extrapolation.

To get a better idea of how much memory the job actually needs at a given point, we employ aggressive garbage collection via Java Virtual Machine (JVM) parameters.
Further, for now, we discount the base level of memory use that is allocated by the framework and the operating system.
The sample runs can and for efficiency purposes should be executed on a single machine, which can be a node of the target infrastructure, but also, for instance, a developer's laptop or desktop.
This is because the conversion of input dataset components to Java objects in memory should be identical as long as the software is comparable, and that is the measure we are interested in.

\begin{figure}[hb]

    \hspace{0mm}
    \subfloat{%
        \includegraphics[width=.6\linewidth]{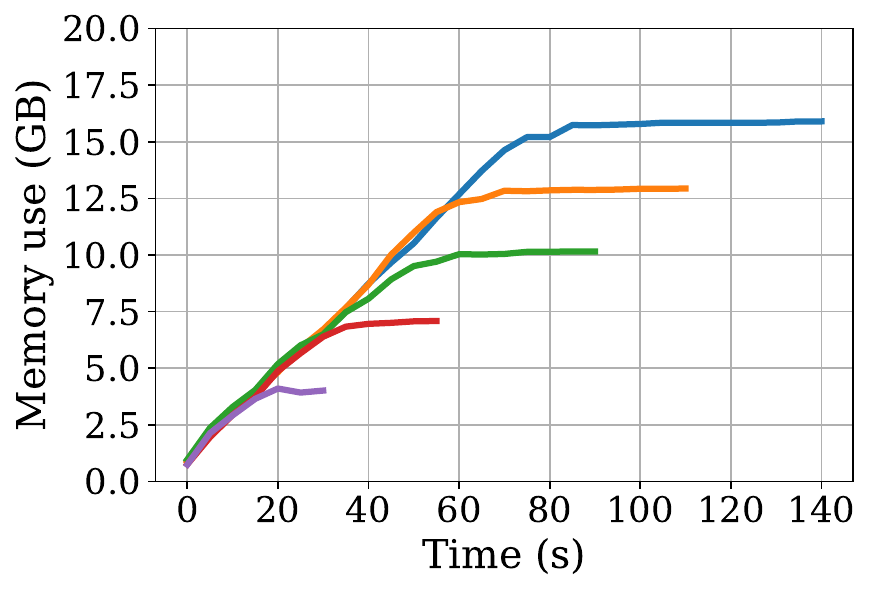}
    }
    \subfloat{%
        \hspace{-1.2mm}
        \includegraphics[width=.30\linewidth]{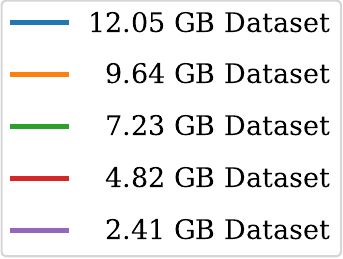}
        \vspace{10.2mm}
    }
    \caption{Memory use on a single-node machine measured over time for K-means on Spark, executed with five linearly distributed dataset sample sizes.}\label{fig:profiling}
\end{figure}

Figure~\ref{fig:profiling} shows an example of the job profiling and memory usage monitoring for K-Means on Spark.
In this case, the job's memory usage grows linearly with the input dataset size and can thus be confidently extrapolated to larger input dataset sizes.

\subsection{Categorization of Jobs Based on Memory Usage Patterns}\label{ssec:categorization}

After gathering all memory usage data from the profiling runs, we need to interpret it, so that we can use this information to accelerate the search for optimal resources for a given job.
Our goal is to model the relation between input data size and used memory during job execution in order to estimate the required amount of memory for resource-efficient processing.
We identified three main scenarios regarding the readings from the memory profiling, resulting in the following three categories, whereby the second one is an expansion of Crispy by Ruya.
\\[1.5ex]\textbf{1.) Linear:}\\
In this case, there is a linear relationship between memory use and input dataset size.
This is typically the case for iterative jobs, which load the whole dataset into memory at once and keep it throughout the execution.
\\[1.5ex]\textbf{2.) Flat:}\\
Here, there is no apparent correlation between memory use and input dataset size, which means that the memory use remains flat as the input dataset size increases.
This can be due to the job or the framework itself not making use of the node's memory capacity.
Examples are jobs based on one-pass algorithms, like Grep, or jobs running on a distributed dataflow framework like Hadoop, which writes all data to disk between the stages of a distributed dataflow job.
\\[8.5ex]\textbf{3.) Unclear:}\\
Another possibility is that there is no linear correlation between input dataset size and measured memory use, but they are still somehow correlated.
This can be the case for iterative jobs, which operate on the whole dataset or large parts of it at once and continuously generate new Java objects faster than the garbage collection can keep up.
As such, the memory use will increase as the execution time goes on, however, this does typically not happen linearly, due to periodic garbage collection.
Another possibility is that a job's memory requirement grows quicker or slower than linearly in relation to the input dataset, e.g., when the memory use grows exponentially or logarithmically.
However, this does not appear to be a common case for big data jobs.
\\[1.0ex]
A simple way of categorizing the readings from the monitoring process is achieved through training a linear regression model on the data.
In case we observe a high accuracy on the training data itself, e.g., a score of $>.99$ when using the $R^2$ scoring metric, we assume the relation to be linear.
Then, the trained model can be used to estimate memory requirements for the actual production jobs, depending on the size of their input dataset.
In case of a low $R^2$ score, e.g. $<0.1$, we consider the relation between the input dataset size and the memory use to be uncorrelated, and therefore the memory needs of the job to be flat.
Consequently, we categorize a job profiling run with a resulting model's $R^2$ score between 0.1 and 0.99 as unclear.

\subsection{Splitting the Resource Configuration Search Space}

To execute this job on the full dataset, we need to select a suitable cluster resource configuration consisting of a node type and a scale-out.
The node types differ primarily in their number of cores and RAM per core, but especially in public clouds, some nodes specialize in network speed, I/O speed and may be equipped with a GPU or a CPU based on a different architecture like ARM\@.
Our main goal is a drastic reduction of the search space by temporarily excluding the configurations that are conflicting with the job's memory requirement.

We get the final requirement of total cluster memory by combining the memory requirement of the job itself with the overhead by the operating system and the distributed dataflow framework.
Here, it is also appropriate to add to the memory requirement as leeway to account for slight miscalculations or even phenomena like different Java Virtual Machine (JVM) implementations leading to slightly different footprints of objects in the JVM heap.

For jobs with a \emph{flat} memory requirement, we limit the prioritized search space to configurations with comparatively low total memory, since for these jobs, additional memory only adds to the cost without improving the performance.
When choosing the exact size of the priority group, there is a trade-off between larger groups leading to longer search and smaller groups running the risk of not including the optimal configuration.
We propose 10\% to 20\%, whereby the lower percentage values should be more promising for larger search spaces and higher ones for smaller search spaces.

For jobs with \emph{unclear} memory requirements, we cannot limit the search space, which results in unmodified Bayesian optimization.

For jobs with memory requirement that grows \emph{linearly} with the input dataset size, we can prioritize jobs with at least as much total cluster memory as per requirement.
In case excess memory leads to a significant efficiency decrease, the process would quickly converge on the low-memory configurations.
If the memory requirement is determined to be too high to be fulfilled by any of the available cluster configurations, we prioritize configurations with very high or very low total cluster memory, because some jobs can make use of all memory they are given and others need either enough or none.
Depending on the job and its implementation of caching, the job may keep a significant part of the dataset in memory and read the rest from disk on demand, or, in the worst case, spill each object to and read back from the disk on every iteration of the job's underlying algorithm.

\subsection{Bayesian-Optimized Iterative Search with CherryPick}

For navigating the search space and thereby iteratively choosing new configurations to try, we employ Bayesian optimization, as described in \emph{CherryPick}~\cite{cherrypick}.
However, our approach should also work with any other implementation of iterative search via Bayesian optimization.
Here, each configuration is encoded by its principal features like the number of cores and the amount of memory.
The idea here is to try three initial configurations randomly, observing the resulting costs.
For the rest of the unexplored search space, the posterior distribution is estimated based on the already available data points.
For each subsequent iteration, the next configuration is chosen based on an estimated function called \emph{acquisition function}, with the most prominent including ``probability of improvement'' and ``expected improvement''.
Like CherryPick, we employ the latter, which chooses the next configuration that is believed to yield the most significant cost savings compared to the best previously tried configuration.

We limit the initial search space by only considering configurations that comply with the previously determined total cluster memory requirement.
Then, we start the exploration of this reduced search space via Bayesian optimization.
Only after exhaustively examining the search space consisting of prioritized configurations, we start to explore the search space with the remaining configurations, utilizing the knowledge gained from the previous search as a starting point.

Without having tried every resource configuration, it is not possible to know when one has found the optimal one.
In general, the search process ends when the expected improvement does not justify the potential cost of an execution on a configuration that is worse than the best out of the previously seen ones.
However, the savings from a more efficient resource configuration are cumulated over each subsequent iteration.
Therefore, the main determining factor of whether one can justify further search is the expected number of future job executions.

\section{Evaluation}\label{sec:EVALUATION}
This section contains the evaluation of the Ruya approach\footnote{\href{https://github.com/dos-group/ruya}{github.com/dos-group/ruya}} to iterative cluster configuration.
Specifically, the quality of the resource selections and the profiling time overhead are evaluated.

\vspace{3mm}
\subsection{Experimental Setup}\label{ssec:experimental_setup}

The profiling phase via Crispy\footnote{\href{https://github.com/dos-group/crispy}{github.com/dos-group/crispy}}, was conducted on one co-author's work laptop, which is a 2020 T14 Thinkpad with 32~GB RAM and an AMD Ryzen 7 PRO 4750U CPU (up to 4.10~GHz).
This is mainly relevant for the profiling speed.%

\vspace{1mm}
\subsubsection{Prototype Implementation}

For the prototype implementation of the Ruya cluster configuration method, we chose Python (version 3.10) for its wealth of available libraries and  its code readability.
One such library, in particular, is \textit{Scikit-Learn} (v. 1.0.2)~\cite{scikit-learn}, which we benefited from when building the memory usage model and implementing the Gaussian process on which the Bayesian optimization depends.
Major supporting libraries we used were \textit{Numpy} (v. 1.22.0) and \textit{Pandas} (v. 1.3.5).
The local profiling experiments ran on Java 8, Spark 2.1.1, and Hadoop 2.7.3.

\vspace{1mm}
\subsubsection{Datasets}

For the evaluation of our methods, we used an existing dataset\footnote{\href{https://github.com/oxhead/scout}{github.com/oxhead/scout}, accessed August 2022} which was introduced by Hsu et al.\ in \emph{Arrow}~\cite{hsu2018arrow}.
The dataset contains 1031 unique Spark and Hadoop executions, which were facilitated by the benchmarking tool \emph{HiBench} by Intel, and which ran on 69 different AWS cluster configurations.
There are seven different underlying algorithms and each job was executed with two different dataset sizes, with the smaller one being called ``huge'' while the larger one is called ``bigdata''.

The cluster configurations have scale-outs between 4 and 48 machines and they have machine types of classes \emph{c}, \emph{m}, and \emph{r} in sizes \emph{large}, \emph{xlarge}, and \emph{2xlarge}.
Virtual machines of the \emph{c} type have less memory per core than those of the type \emph{r}, while machines of the \emph{m} type lie between those two.
Denominations like \emph{large}, \emph{xlarge}, and \emph{2xlarge} refer to the number of cores per machine.

\vspace{3mm}

\subsection{Memory Usage Measurement and Modeling}
\vspace{1mm}

Since the distributed data flow frameworks that we examine run on the JVM, they are subject to garbage collection.
To get a good idea of how much memory is actually being used, we need to make sure that unused objects are deleted from memory without substantial delay.
As detailed in Crispy~\cite{will2022get}, we achieved this by configuring the JVM to use more aggressive garbage collection.
This allows for more accurate memory usage readings during profiling runs at the expense of reasonably longer runtimes.

\begin{table}[ht]
    \centering
    \caption{Determined Job Memory Requirement}
    \vspace{-1mm}
    \label{tab:memory_requirements}
    \begin{tabular}{lll|llrl}
        &&&& \multicolumn{3}{c}{Result\hspace{69mm}}\\
        \hline \\[-1.7ex]
        Naive Bayes & Spark  & bigdata & \hspace{3mm} &  linear: & \hspace{-4mm} 754 & \hspace{-4mm} GB \\
        Naive Bayes & Spark  & huge    &              &  linear: & \hspace{-4mm} 395 & \hspace{-4mm} GB \\\rowcolor{Gray}
        K-Means     & Spark  & bigdata &              &  linear: & \hspace{-4mm} 503 & \hspace{-4mm} GB \\\rowcolor{Gray}
        K-Means     & Spark  & huge	   &              &  linear: & \hspace{-4mm} 252 & \hspace{-4mm} GB \\
        Page Rank   & Spark  & bigdata &              &  linear: & \hspace{-4mm}  86 & \hspace{-4mm} GB \\
        Page Rank   & Spark  & huge    &              &  linear: & \hspace{-4mm}  42 & \hspace{-4mm} GB \\\rowcolor{Gray}
        Log. Regr.  & Spark  & bigdata &              & unclear  &                    &                  \\\rowcolor{Gray}
        Log. Regr.  & Spark  & huge	   &              & unclear  &                    &                  \\
        Lin. Regr.  & Spark  & bigdata &              & unclear  &                    &                  \\
        Lin. Regr.  & Spark  & huge	   &              & unclear  &                    &                  \\\rowcolor{Gray}
        Join        & Spark  & bigdata &              &    flat  &                    &                  \\\rowcolor{Gray}
        Join        & Spark  & huge	   &              &    flat  &                    &                  \\
        Page Rank   & Hadoop & bigdata &              &    flat  &                    &                  \\
        Page Rank   & Hadoop & huge    &              &    flat  &                    &                  \\\rowcolor{Gray}
        Terasort    & Hadoop & bigdata &              &    flat  &                    &                  \\\rowcolor{Gray}
        Terasort    & Hadoop & huge    &              &    flat  &                    &                  \\
        \hline
    \end{tabular}
\vspace{-3mm}
\end{table}

In accordance with Subsection~\ref{ssec:categorization} of the approach, we set the boundaries of the categorization to an $R^2$ score of $0.1$ and $0.99$ for determining whether input dataset size and memory use are linearly correlated, uncorrelated, or ambiguous for a particular job.
Out of the mix of jobs in our evaluation dataset, $\frac{6}{16}$ were determined to be linearly scaling, another $\frac{6}{16}$ were found to be static, and the remaining $\frac{4}{16}$ could not be determined due to unclear readings during the memory profiling process.

The exact results of our memory profiling phase are shown in Table~\ref{tab:memory_requirements}.
Here, ``flat'' describes the situation where a job's memory requirement does not scale with the size of the input dataset, ``unclear'' describes an unknown memory requirement, while for linearly scaling memory requirements, our trained linear regression model provided an estimate of the required memory for the job itself in GB, excluding overhead by the framework and the operating system which still has to be considered for each cluster node.

\begin{table*}[t!]
    \centering
    \caption{The Results in Detail: A Configuration with the Normalized Cost \textbf{\emph{c}} is Found After How Many Iterations?}
    \begin{tabular}{l@{\hskip 1.9mm }l@{\hskip 1.98mm }l@{\hskip 4mm}l|rrr|rrr|@{\hskip 3mm}|rrr}
        \multicolumn{4}{c}{} & \multicolumn{3}{c}{\normalsize\textbf{CherryPick}}  & \multicolumn{3}{c}{\normalsize\textbf{Ruya}} & \multicolumn{3}{c}{\normalsize Quotient \large $\frac{\text{Ruya}}{\text{CherryPick}}$} \\[1.4ex]
        &  &  &  &
        \textbf{c\scalebox{.7}{\,$\le$\,}1.2} &
        \textbf{c\scalebox{.7}{\,$\le$\,}1.1} &
        \textbf{c\scalebox{.7}{\,$=$\,}1.0} &
        \textbf{c\scalebox{.7}{\,$\le$\,}1.2} &
        \textbf{c\scalebox{.7}{\,$\le$\,}1.1} &
        \textbf{c\scalebox{.7}{\,$=$\,}1.0} &
        \textbf{c\scalebox{.7}{\,$\le$\,}1.2} &
        \textbf{c\scalebox{.7}{\,$\le$\,}1.1} &
        \textbf{c\scalebox{.7}{\,$=$\,}1.0} \\
        \hline\\[-1.7ex]
        Naive Bayes & Spark  & bigdata & (linear)  &  1.775  &   4.840  &  44.510 &   1.835 &  5.465 & 44.505   &     103.4\% &     112.9\% &      100.0\%  \\
        Naive Bayes & Spark  & huge    & (linear)  &  4.075  &   9.485  &  18.385 &   1.315 &  1.675 &  4.190   & \gn{32.3}\% & \gn{17.7}\% &  \gn{22.8}\% \\\rowcolor{Gray}
        K-Means     & Spark  & bigdata & (linear)  & 10.750  &  17.320  &  17.320 &   2.440 &  4.355 &  4.355   & \gn{22.7}\% & \gn{25.1}\% &  \gn{25.1}\% \\\rowcolor{Gray}
        K-Means     & Spark  & huge	   & (linear)  &  6.665  &  13.065  &  15.345 &   3.175 &  7.470 &  7.470   & \gn{47.6}\% & \gn{57.2}\% &  \gn{48.7}\% \\
        Page Rank   & Spark  & bigdata & (linear)  &  4.840  &  18.935  &  18.935 &   4.680 & 15.535 & 15.535   & \gn{96.7}\% & \gn{82.0}\% &  \gn{82.0}\% \\
        Page Rank   & Spark  & huge    & (linear)  &  4.870  &  11.440  &  17.570 &   5.195 & 12.230 & 17.210   &     106.7\% &     106.9\% &       98.0\%  \\\rowcolor{Gray}
        Lin. Regr.  & Spark  & bigdata & (unclear) &  2.675  &   7.040  &  25.870 &   2.685 &  7.125 & 22.615   &     100.4\% &     101.2\% &       87.4\%  \\\rowcolor{Gray}
        Lin. Regr.  & Spark  & huge	   & (unclear) &  7.760  &  10.470  &  14.275 &   6.770 &  9.790 & 13.165   &      87.2\% &      93.5\% &       92.2\%  \\
        Log. Regr.  & Spark  & bigdata & (unclear) &  7.550  &  10.820  &  10.820 &   7.840 & 11.205 & 11.205   &     103.8\% &     103.6\% &      103.6\%  \\
        Log. Regr.  & Spark  & huge	   & (unclear) &  6.375  &  10.575  &  11.620 &   5.925 & 10.095 & 11.015   &      92.9\% &      95.5\% &       94.8\%  \\\rowcolor{Gray}
        Join        & Spark  & bigdata & (flat)    &  7.200  &  18.180  &  32.935 &   1.000 &  1.800 &  5.515   & \gn{13.9}\% & \gn{ 9.9}\% &  \gn{16.7}\% \\\rowcolor{Gray}
        Join        & Spark  & huge	   & (flat)    &  9.370  &  16.825  &  22.800 &   1.215 &  2.770 &  6.515   & \gn{13.0}\% & \gn{16.5}\% &  \gn{28.6}\% \\
        PageRank    & Hadoop & bigdata & (flat)    &  7.500  &  27.510  &  35.025 &   1.220 &  2.750 &  5.990   & \gn{16.3}\% & \gn{10.0}\% &  \gn{17.1}\% \\
        PageRank    & Hadoop & huge    & (flat)    & 12.320  &  15.285  &  20.645 &   1.345 &  2.705 &  5.750   & \gn{10.9}\% & \gn{17.7}\% &  \gn{27.9}\% \\\rowcolor{Gray}
        Terasort    & Hadoop & bigdata & (flat)    & 13.530  &  35.475  &  35.475 &   1.965 &  5.165 &  5.165   & \gn{14.5}\% & \gn{14.6}\% &  \gn{14.6}\% \\\rowcolor{Gray}
        Terasort    & Hadoop & huge    & (flat)    & 32.505  &  36.535  &  36.535 &   4.305 &  5.890 &  5.890   & \gn{13.2}\% & \gn{16.1}\% &  \gn{16.1}\% \\
        \hline
        \hline\\[-1.7ex]
        Mean        &        &         &         &  8.735  &  16.487  &  23.629 &   3.307 &  6.627 & 11.631   & \gn{37.9\%}  & \gn{40.2\%} & \gn{49.2\%} \\
    \end{tabular}\label{tab:details}
\end{table*}

\subsection{Configuration Selection}

To evaluate our selection strategy, we compare it to Cherry\-Pick as a baseline, which we implemented based on the description in the corresponding paper~\cite{cherrypick}.

We specifically investigate the monetary cost, since in public clouds like AWS, this is an adequate indicator of resource-efficiency.
For a job, the cost of each cluster configuration is normalized to the configuration that had the cheapest execution.
This means that the cheapest cluster configuration for a job always has a cost of 1.0, and another configuration that resulted in a three and a half times higher execution cost for this job would then have a cost of 3.5.

Table~\ref{tab:details} shows the results in detail, as we compare the baseline approach CherryPick with Ruya.
In particular, we examine how many iterations the search process needs to find the optimal or a near-optimal configuration, with near-optimal configurations being up to 10\% or 20\% more costly than the optimal one.
Our results are averaged over 200 experiment iterations to account for variance due to the random initialization at the beginning of the Bayesian optimization process.

\pagebreak
The improvement to the baseline for each job correlates with the degree to which the search space had been reduced.
Thus, we see that for the linear and logistic regression jobs the results are equal to the baseline, since modeling of memory usage was forgone here due to their classification as unclear.
For jobs with flat memory requirements on the other hand, notably Hadoop jobs, choosing promising configurations was rather straightforward.
Here, the ten configurations with the lowest total memory were put in the priority group, which accounts for around $\frac{1}{7}$ of the total search space in our evaluation dataset.
For jobs with a classification as linear, we generally see a reduction in search operations due to a reduction of the search space.
However, for PageRank on Spark with the ``huge''-sized dataset, this memory requirement was so low that all configurations fulfilled it, leading to no search space reduction.
Another exception is Naive Bayes on Spark for the ``bigdata''-sized dataset, where Ruya misclassified two configurations as fulfilling the memory requirement due to an inaccurate extrapolation, even though actually none of the available configurations have enough total memory.

Averaged across all jobs, we see that Ruya requires less than half the amount of iterations to find the optimal resource configuration as compared to CherryPick, with this figure moving further in Ruya's favor for near-optimal configurations.

\begin{figure}[htb]
    \centering
    \includegraphics[width=.77\linewidth]{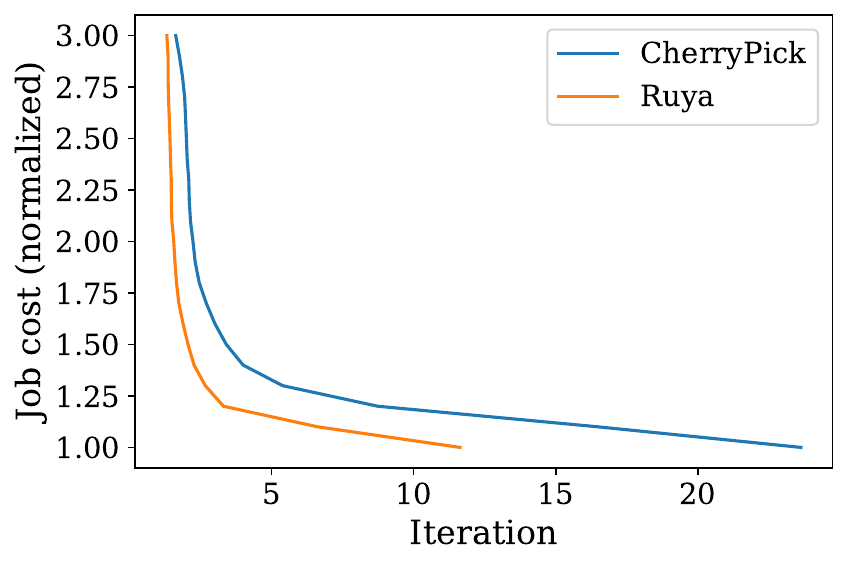}
    \vspace{-1mm}
    \caption{Normalized cost of a job execution with the best discovered configuration at a given iteration, averaged over all jobs.}\label{fig:job_cost}
    \vspace{-2mm}
\end{figure}

Figure~\ref{fig:job_cost}, summarizes the performance of CherryPick and Ruya averaged over all evaluated jobs.
It shows for each iteration the normalized cost of the best found configuration up until that point.
On average, Ruya needs around 12 iterations to find the optimal resource configuration, while CherryPick needs around 24, while for less-than-optimal configurations this difference in search speed is even more pronounced.

\begin{figure}[htb]
    \centering
    \includegraphics[width=.77\linewidth]{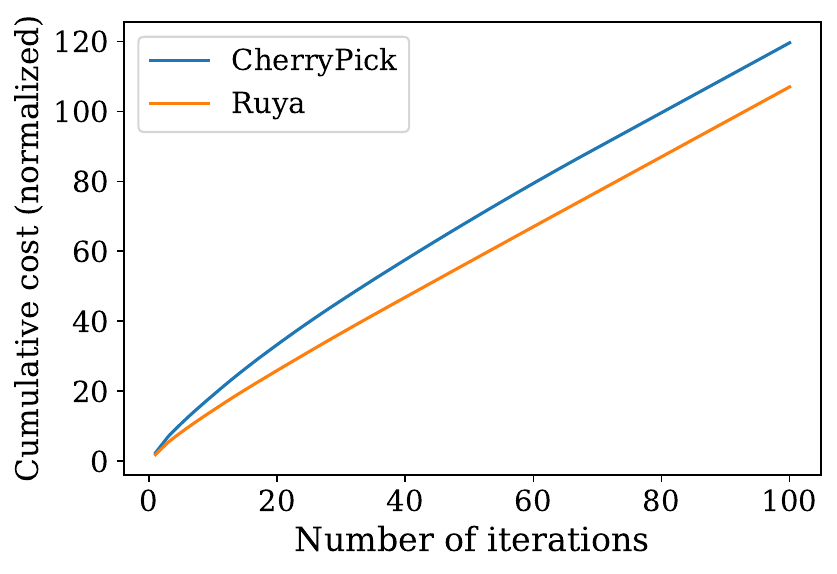}
    \vspace{-1mm}
    \caption{Comparison of Ruya and CherryPick over multiple iterations of job executions, averaged over all jobs.}\label{fig:cumulative_job_cost}
\end{figure}

Figure~\ref{fig:cumulative_job_cost} shows the cumulation of the normalized execution costs over the iterations of this process.
We see that the relative difference is most pronounced for a number of executions that is below 25 and then starts to lose significance as more iterations happen.
However, as the number of recurrences grows, at some point there may be enough performance metrics available to employ complex performance models to estimate runtimes and suggest configurations, which also account for the runtime-cost trade-off.
The exact values of this comparison depend on the exact implementation of the stopping criterion.
Further, this figure assumes that the input data size and therefore computational effort remains equal throughout the iterations.

\subsection{Profiling Speed}

Due to Ruya's need for memory profiling runs, there is some time overhead before the first full job execution.

Table~\ref{tab:profiling_times} shows that profiling times were between two and 20 minutes, while the average was just below ten minutes.
The median is below eight minutes on the hardware as described in Subsection~\ref{ssec:experimental_setup}.

\begin{table}[htb]
    \centering
    \caption{Memory Profiling Time for all Jobs}
    \label{tab:profiling_times}
    \vspace{-1mm}
    \begin{tabular}{l@{\hskip 1.3mm }l@{\hskip 1.98mm }l|cr}
        && $\;\;\;\;\;\;\;$ && {\hspace{3mm} Time (s)\hspace{-2mm}} \\
        \hline\\[-1.7ex]
            Naive Bayes & Spark  & bigdata & \hspace{5mm} & 373  \\
            Naive Bayes & Spark  & huge    &     & 369  \\\rowcolor{Gray}
            K-Means     & Spark  & bigdata &     & 470  \\\rowcolor{Gray}
            K-Means     & Spark  & huge	   &     & 470  \\
            Page Rank   & Spark  & bigdata &     & 1292 \\
            Page Rank   & Spark  & huge	   &     & 1292 \\\rowcolor{Gray}
            Lin. Regr.  & Spark  & bigdata &     & 372  \\\rowcolor{Gray}
            Lin. Regr.  & Spark  & huge	   &     & 198  \\
            Log. Regr.  & Spark  & bigdata &     & 675  \\
            Log. Regr.  & Spark  & huge	   &     & 562  \\\rowcolor{Gray}
            Join        & Spark  & bigdata &     & 136  \\\rowcolor{Gray}
            Join        & Spark  & huge    &     & 110  \\
            Page Rank   & Hadoop & bigdata &     & 812  \\
            Page Rank   & Hadoop & huge    &     & 812  \\\rowcolor{Gray}
            Terasort    & Hadoop & bigdata &     & 547  \\\rowcolor{Gray}
            Terasort    & Hadoop & huge    &     & 547  \\
        \hline\hline\\[-1.7ex]
            Mean        &        &         &     & 565  \\
    \end{tabular}
\end{table}

In general, this profiling overhead is not significant, even for a single job execution on a large dataset and becomes less relevant as more search iterations are performed.
The profiling process only needs to be repeated if some aspects of the execution context changes, such as job parameters or key dataset characteristics other than just the size, like formatting.

Also, the profiling overhead is irrespective of the size of the full dataset.
This is because the dataset sample sizes are not chosen as a fixed portion of the original dataset, but iteratively to achieve just a long enough runtime that allows for reliable memory usage measurements.

\subsection{Discussion}

Overall, the evaluation results indicate that, on average, Ruya provides a significant improvement over the baseline, as it needs less than half as many iterations to find optimal and near-optimal configurations.
Ruya achieves this without knowledge from prior executions of the same job, incurring only small resource and time cost expenditures for executing the profiling runs.
Ten minutes of profiling on a personal computer is essentially free of cost, and the extent of this profiling effort is irrespective of the size of the full dataset.
In cases where the memory usage modeling and extrapolation does not work, e.g., the memory readings are not accurate enough, Ruya is largely able to recognize this and falls back to the baseline approach, which is regular Bayesian optimization.
Therefore, Ruya has shown to be about as good or better than the baseline approach for each of the 16 jobs in our evaluation.
This improvement is most pronounced in earlier iterations and loses significance once more iterations are made.
However, once more information from previous iterations becomes available, resource selection methods that employ performance models would become increasingly viable.

Another key advantage of Ruya is that it considers the input dataset size in the resource configuration selection, whereas a system like CherryPick would effectively need to restart the profiling process once these key input dataset characteristics change.
This allows it to react effectively to changes in dataset size, which can be expected in real world applications due to changing and especially growing datasets with data that is continuously being generated.

\section{Related Work}\label{sec:RELATED_WORK}

Ruya is a black-box approach, meaning that it aims to be agnostic to specific data processing systems.
Thus, the first two subsections discuss the two main categories of black-box approaches to configuring clusters for recurring data analytics jobs.
Lastly, we also describe related work that focuses on different aspects of memory-aware large-scale data processing.

\subsection{Approaches Based on Historical Performance Data}

Some approaches use runtime data to predict the job's scale-out and runtime behavior.
This data is gained either from dedicated profiling or previous full executions~\cite{ernest, hongzi2016resource, chen2021silhouette, rajan2016perforator, bell, scheinert2021bellamy, will2021c3o, scheinert2021potential}.
The models can then be used to predict the execution performance for different cluster configurations, and the most resource-efficient one will be chosen.
This can, for example, be the one with the lowest expected execution cost in a public cloud scenario under consideration of potential additional constraints.

\emph{Ernest}~\cite{ernest} trains a parametric model for the scale-out behavior of jobs on the results of sample runs on reduced input data, which works well for recurring programs that exhibit a rather intuitive scale-out behavior.
Initial configurations are tried out based on optimal experiment design.

With \emph{C3O}~\cite{will2021c3o}, the execution context is taken into consideration, and the effects of performance-influencing factors other than the cluster setup are modeled.
This enables learning from previous job executions even if they had vastly different parameter and dataset inputs than the current job.

\emph{Silhouette}~\cite{chen2021silhouette} is a cloud configuration selection framework based on performance modeling with minimal training overhead.
Silhouette uses advanced statistical techniques and proposes a model transformer for quick transfer learning, and it effectively optimizes cloud configurations under constraints.

The disadvantage of all resource selection approaches based on performance models is that they assume the availability of training data.
The complexity of the models, and thereby the need for larger amounts of training data, grows as more contextual information about previous job executions under slightly different circumstances needs to be incorporated.

In contrast to these approaches, Ruya does not require any performance metrics from previous job executions.
It can function in a cold-start scenario for a job that does not yet have performance metrics from previous executions.
Ruya purposefully collects all performance metrics for the job at hand, while the costs of initial iterations are amortized by efficiency gains in all subsequent recurrences of the job.

\subsection{Approaches Based on Iterative Search}

More closely related approaches configure the cluster iteratively, attempting to find a better configuration at each iteration by utilizing runtime information from prior iterations.
There are different strategies for reaching a stopping criterion, i.e., when it is expected that further exploration of the search space will not lead to significant enough exploitation potential to justify the additional search overhead~\cite{cherrypick, hsu2018micky, hsu2018arrow, fekry2020accelerating, he2019statistics}.

\emph{CherryPick}~\cite{cherrypick} tries to directly predict the optimal cluster configuration that obeys given runtime targets by utilizing Bayesian optimization.
The search stops once it has found the optimal configuration with reasonable confidence.

In \emph{Arrow}~\cite{hsu2018arrow}, the authors use low-level metrics such as memory utilization to enhance this aforementioned method.
These metrics are, for example, CPU utilization, working memory size, and I/O wait time.
Utilizing these allows for finding optimal or near-optimal configurations with fewer iterations.
The authors also compared Arrow to CherryPick as a baseline.
Unlike Ruya, their results appear to only slightly improve the speed of finding optimal and near-optimal solutions compared to CherryPick.

\emph{SimTune}~\cite{fekry2020accelerating} combines workload characterization and multi-task Bayesian optimization to speed up its incremental search for near-optimal configurations.
The idea is to find similarities with previous workloads and thereby infer information about the resource usage patterns of the current job.
Its setup allows the tuning to be done online.
Compared to SimTune, Ruya is suitable for cold-start scenarios since it does not assume that there have been similar prior executions of a job and it learns memory usage patterns through quick profiling runs on reduced hardware and small input dataset samples.

Overall, these highlighted methods use Bayesian optimization and typically enhance this process by incorporating further information about the job's resource usage patterns into the search, but they disregard the occurrence and relevance of memory bottlenecks.
Ruya finds good configurations quickly by reducing the search space to prioritize configurations which avoid prohibitively costly memory bottlenecks, based on information gained through efficient and inexpensive profiling runs.

\subsection{Memory Management and Memory Allocation Methods}

This subsection describes related work that focuses on different aspects of memory-aware large-scale data processing.

In \emph{Crispy}~\cite{will2022get}, we presented an approach to choosing a suitable cluster configuration for unique, one-off jobs.
We recognized the significance of adequate cluster memory allocation for resource efficiency and introduced the idea of profiling a job for its memory access patterns, measured through APIs on the operating system level.
Based on estimated memory needs, we then execute the job on a cluster configuration that is expected to perform best.
In Ruya, we utilized the profiling technique and extended the memory usage estimation of this approach.

\emph{Blink}~\cite{al2022blink} follows a similar approach to Crispy, but is focusing on Apache Spark.
Here, the memory footprint of the cached dataset samples is continuously extracted through the SparkListener API, which allows for accurate readings.
In contrast, our approaches are in principle independent of a specific dataflow framework by measuring memory use on the operating system level.
However, when only regarding Spark jobs, Blink could be used in combination with Ruya as a replacement for Crispy.

\emph{Apache Flink}~\cite{flink} provides an API specifically for iterative computations like some of the iterative machine learning jobs that were evaluated in this paper~\cite{mior2020respark}.
It manages caching in a way that in case of a lack of memory, as much as possible of the dataset is enduringly cached, while the remaining part is repeatedly read from disk on demand.
In iterative algorithms, where in each iteration the whole dataset is processed, this avoids spilling the least recently used objects to disk, which would ultimately lead to reading all objects from disk at each iteration.
This approach can lower the impact of memory bottlenecks, but avoiding them by allocating enough memory to a job is to be preferred.

\emph{Juggler}~\cite{al2022juggler} has a similar objective but targets Apache Spark.
It extends Spark by automating the selection of datasets and parts of datasets to cache, not just for jobs based on iterative algorithms, in addition to selecting cluster resources based on the previous step.
In contrast, Ruya focuses on just selecting resources for already existing jobs, which it treats as a black box, and does not attempt to assist users in the job design process.
Therefore Ruya is not dependent on any specific distributed dataflow framework.

\section{Conclusion}\label{sec:CONCLUSION}
This paper presented Ruya, a system that profiles a distributed dataflow job on a single machine and uses the knowledge about the job's memory requirements to efficiently search for optimized cluster configurations, avoiding costly memory bottlenecks.

In our experimental evaluation, we see Ruya successfully modeling memory use in relation to input dataset size for the majority of examined jobs.
For the other cases, where our current methods of measuring and modeling memory consumption fall short, we show that Ruya reliably falls back to a baseline approach.
In total, there is a reduction of required search iterations to find optimal and near-optimal configurations by over 50\% from our baseline.
Ruya spent an average of less than ten minutes for job profiling runs on a consumer-grade laptop.

Our approach can adapt to changing input dataset sizes without having to restart the search process for a suitable cluster configuration.

Future work may include managing and sharing knowledge gained from profiling and iteratively searching as more and more performance data becomes available.
This information could then be used to train complex performance models that consider the wider execution context when assisting the user in choosing suitable cluster resources.

\section*{Acknowledgments}

This work has been supported through grants by the German Ministry for Education and Research (BMBF) as BIFOLD (grant 01IS18025A) and by the German Research Foundation (DFG) as FONDA (DFG Collaborative Research Center 1404).

\bibliographystyle{IEEEtran}
\balance
\bibliography{./references}

\end{document}